\documentclass{PoS}
\input{psfig}
\def\gsim{ \lower .75ex \hbox{$\sim$} \llap{\raise .27ex \hbox{$>$}} }
\def\lsim{ \lower .75ex\hbox{$\sim$} \llap{\raise .27ex \hbox{$<$}} }

\def\beq{\begin{equation}}
\def\eeq{\end{equation}}

\def\sw{{\it Swift}}

\def\ep{$E_{\rm p}$}

\def\epo{$E^{\rm obs}_{\rm p}$}

\def\liso{$L_{\rm iso}$}
\def\eiso{$E_{\rm iso}$}
\def\ama{$E_{\rm p}-E_{\rm iso}$}
\def\yone{$E_{\rm p}-L_{\rm iso}$}
\def\ghi{$E_{\rm p}-E_{\gamma}$}

\def\th{$\theta_{\rm jet}$}

\def\egamma{$E_{\gamma}$}

\def\G{$\Gamma_{0}$}

\def\lisocom{$L'_{\rm iso}$}
\def\eisocom{$E'_{\rm iso}$}
\def\egcom{$E'_{\gamma}$}

\def\epcom{$E'_{\rm p}$}

\title{GRBs have preferred jet opening angles and bulk Lorentz factors
}

\ShortTitle{$\theta_{\rm jet}$ and $\Gamma_0$ of GRBs}

\author{\speaker{ G. Ghisellini$^1$}\thanks{E-mail: gabriele.ghisellini@brera.inaf.it}, 
 G. Ghirlanda$^1$, R. Salvaterra$^2$, L. Nava$^3$, D. Burlon$^4$, G. Tagliaferri$^1$,  
  S. Campana$^1$, S. Covino$^1$, P. D'Avanzo$^1$, A. Melandri$^1$ \\
        1: INAF -- Osservatorio Astronomico di Brera, Italy\\
        2: INAF -- IASF Milano, Italy\\
        3: APC Universit\'e Paris Diderot,  F-75205 Paris Cedex 13, France \\
        4: Sydney Institute for Astronomy, School of Physics, The University of Sydney, NSW 2006, Australia\\
        }
\abstract{
We recently found that Gamma--Ray Burst energies and luminosities, in their comoving frame,
are remarkably similar.
This, coupled with the clustering of energetics once corrected for the collimation
factor, suggests the possibility that all bursts, in their comoving frame,
have the same peak energy \epcom\ (of the order of a few keV) and the same energetics
of the prompt emission \egcom\ (of the order of $2\times 10^{48}$ erg).
The large diversity of bursts energies is then due to the different bulk Lorentz factor \G\
and jet aperture angle \th.
We investigated, through a population synthesis code, what are the 
distributions of \G\ and \th\ compatible with the observations.
Both quantities must have preferred values, with log--normal best fitting distributions
and  $\langle\Gamma_0 \rangle\sim 275$ and $\langle \theta_{\rm jet}\rangle\sim 8.7^\circ$. 
Moreover, the peak values of the \G\ and \th\ distributions must be related --
$\theta_{\rm jet}^{2.5}$\G=const: the narrower the jet angle, the larger the bulk 
Lorentz factor.
We predict that $\sim$6\% of the bursts that point to us should not show any jet break in their 
afterglow light curve since they have $\sin\theta_{\rm jet}<1/$\G. 
Finally, we  estimate that the local rate of 
GRBs is $\sim$0.3\% of all local SNIb/c and $\sim$2.5\% of local hypernovae,
i.e. SNIb/c with broad absorption lines.
}

\FullConference{Gamma-Ray Bursts 2012 Conference -GRB2012,\\
		May 07-11, 2012\\
		Munich, Germany}

\begin{document}

\section{Introduction}

The spectral energy correlations in GRBs are still matter of hot debate. 
The isotropic equivalent energy \eiso\ of the prompt phase of long GRBs
correlates with the rest frame peak $E_{\rm p}$ of the $\nu F_{\nu}$ spectrum  
\cite{amati2002}, \cite{amati2007}: $E_{\rm p}\propto E_{\rm iso}^{0.5}$. 
A similar correlation (obeyed also by short events -- \cite{ghirlanda2009}) exists between 
the isotropic equivalent luminosity \liso\ and \ep\ \cite{yonetoku2004}:
$E_{\rm p}\propto L_{\rm iso}^{0.5}$.

If GRBs emit their radiation within a jet of opening angle \th,  
the true energy \egamma$\simeq$\eiso $\theta_{\rm jet}^2$
can be estimated \cite{frail2001}. 
For $\sim$30 GRBs with known \th, \egamma\  
is tightly correlated with \ep\ \cite{ghirlanda2004}, \cite{ghirlanda2007}.

The presence of outliers of the \ama\ correlation 
\cite{band2005}, \cite{nakar2005}, \cite{shahmoradi2011}
\cite{collazzi2012} and the presence of possible instrumental biases 
\cite{butler2009}, \cite{kocevski2012},
caution about the use of these correlations either for deepening into 
the physics of GRBs or for cosmological purposes.
However, even if instrumental selection effects are present, 
it seems that they cannot produce the correlations we see 
\cite{ghirlanda2008} \cite{nava2008}, \cite{ghirlanda2012a}. 
Moreover, a correlation between \ep\ and \liso\ 
is present within individual GRBs as a function of time \cite{firmani2009},
\cite{ghirlanda2010}, \cite{ghirlanda2011a} \cite{ghirlanda2011b}. 

A new piece of information recently added to the puzzle is that the energetics 
in the comoving frame (i.e. $E_{\rm iso}^\prime$, $L_{\rm iso}^\prime$\ and $E_{\rm peak}^\prime$)
are similar for all GRBs \cite{ghirlanda2012a}.  
For about 30 GRBs we \cite{ghirlanda2012a} found that \eiso(\liso)$\propto\Gamma_0^2$ and \ep$\propto$\G; 
in the comoving frame \eisocom$\sim$3.5$\times10^{51}$ erg, 
\lisocom$\sim$5$\times10^{48}$ erg s$^{-1}$ and \epcom$\sim$6 keV
(see \cite{giannios2012} for a theoretical interpretation).
These results suggest that the \ama\ and \yone\ correlations are a sequence of different \G\ factors.

The comoving true energy \egcom\ turns out to be $\sim 2\times 10^{48}$ erg. 
In \cite{ghirlanda2012a} we argued that to have consistency between the \ghi\ 
and the \ama\ correlations we need $\theta_{\rm jet}^2\Gamma_0$ = constant. 
The distribution of \G\ is centered around \G=65 (130) in the case of a wind (uniform) 
density distribution of the circum--burst medium. 

These new findings prompted us to explore the possibility that the 
\ghi\ and the \ama\ correlations result from all bursts having
the same comoving \egcom\ and \epcom\, but different \G\ and \th.
Specifically, we \cite{ghirlanda2012b}
ask whether \th\ and/or \G\ have preferential values or not,
and if there is a relation between them.
To this aim we have performed extensive numerical simulations, along the
guidelines explained below.

\begin{figure*}
\vskip -0.3 cm
\psfig{file=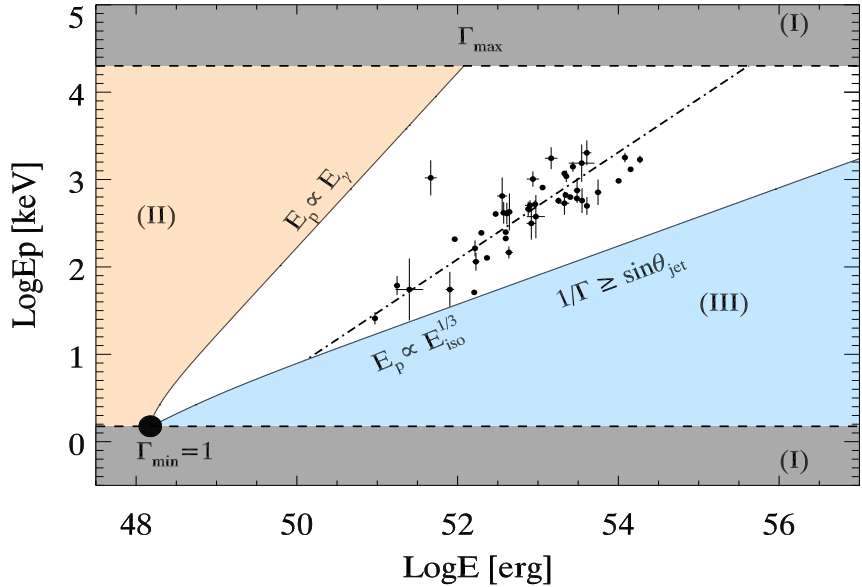,width=14cm,height=9cm}
\vskip -0.5 cm
\caption{
Rest frame plane of GRB energetics. 
The large black dot corresponds to all bursts having the same 
\epcom\ and \egcom. 
For a given \G, the burst moves along the line
$E_{\rm peak}\propto E_\gamma$.
Since we assume $1<\Gamma_0<8000$, regions (I) are forbidden.
Since all our simulated bursts have \th$\leq$90$^\circ$, they cannot lie in region (II). 
For small \G, the beaming cone $\sim 1/$\G\ can become wider than $\theta_{\rm jet}$.
This introduces the limit $E_{\rm peak}\propto E_{\rm iso}^{1/3}$ and bursts cannot lie in
region (III).
Black dots correspond to the real GRBs of the \sw\ complete sample \cite{salvaterra2012}.
}
\label{fg1}
\end{figure*}

\section{Simulation set up}

Fig. \ref{fg1} shows the \ama\ plane. 
The black points are GRBs belonging to the 
complete {\it Swift} sample of \cite{salvaterra2012}.
The large black dot corresponds to our main assumptions, i.e. all bursts,
in the comoving frame, emit \egcom$\sim 2\times 10^{48}$ erg at \epcom$\sim$1.5 keV
{\it independent of their \G.}
\epcom\ is smaller (2$\sigma$) than the mean value derived in \cite{ghirlanda2012a},
in order to be able to reproduce GRBs lying quite close to the $E_{\rm p}\propto E_{\rm iso}^{1/3}$ line.
GRBs with different \G\ would lie on the \ep$\propto$\egamma\ line, giving rise to the \ghi\ 
relation.
Then, by assuming a given aperture angle \th\, we can calculate \eiso.
The GRB will move to the right by the quantity $[1/(1-\cos\theta_{\rm jet})]$ 
if \th$>$1/\G, and by the quantity $2\Gamma_0^2$ otherwise.
In the latter case, the relation between \ep\ and \eiso\ becomes 
$E_{\rm p} \propto E^{1/3}_{\rm iso}$.
This implies that region (III) of Fig. \ref{fg1} is forbidden.
The other forbidden regions are region (II) because this would correspond
to \th$>90^\circ$, and region (I) because we assume 1$<$\G$<$8000.
All our simulated bursts will then lie on the white part of the plane.
The distribution of the simulated bursts in this plane depends on the
chosen distributions of \G\ and \th.
We thus have a tool to find what are the best fitting distributions.

The steps are: 
i) select a redshift from the assumed redshift distribution
(that is taken from \cite{salvaterra2012}, which includes an evolutionary term);
ii) select a \G\ and calculate \ep\ and \egamma; 
iii) select a \th\ and calculate \eiso; 
iv) chose a viewing angle and decide if it is pointing at us or not;
v) calculate the peak flux in the appropriate band (assuming a typical Band spectrum) 
and decide if the burst belongs to the complete \sw\ sample \cite{salvaterra2012} or not. 
Bursts in this sample have a peak flux larger than 2.6 ph cm$^{-2}$ s$^{-1}$,
and almost 90\% of them have a measured redshift.
The steps are repeated until the number of simulated \sw\ bursts matches the real ones.
Finally, we repeat 1,000 times each simulation to see how many times 
we can get a reasonable agreement with several observational constraints.
First, we compare the simulated points of the complete \sw\ sample with the real ones
in the \ama\ plane.
Then we compare them also in the observed planes \epo--Fluence and \epo--Peak Flux
(irrespective if the redshift is known or not).
Finally, we compare the distribution of simulated vs real flux and fluences
of the BATSE and GBM bursts (down to limiting values that are not affected by
incompleteness).

\begin{figure*}
\begin{center}$
\begin{array}{cc}
\hskip -0.7truecm\includegraphics[width=7.8cm,trim=50 20 40 40,clip=true]{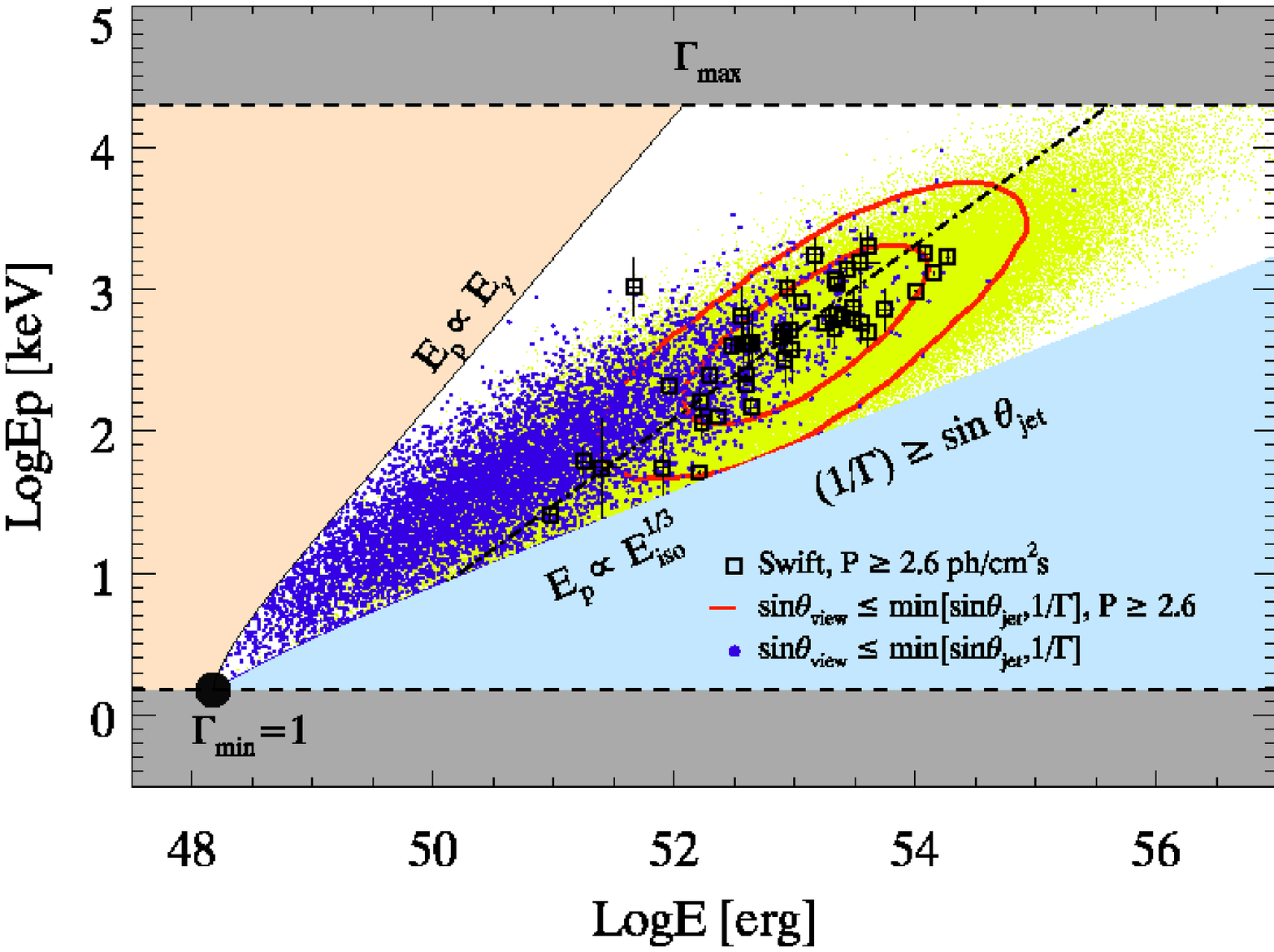} & 
\hskip -0.3truecm\includegraphics[width=7.8cm,trim=50 20 40 40,clip=true]{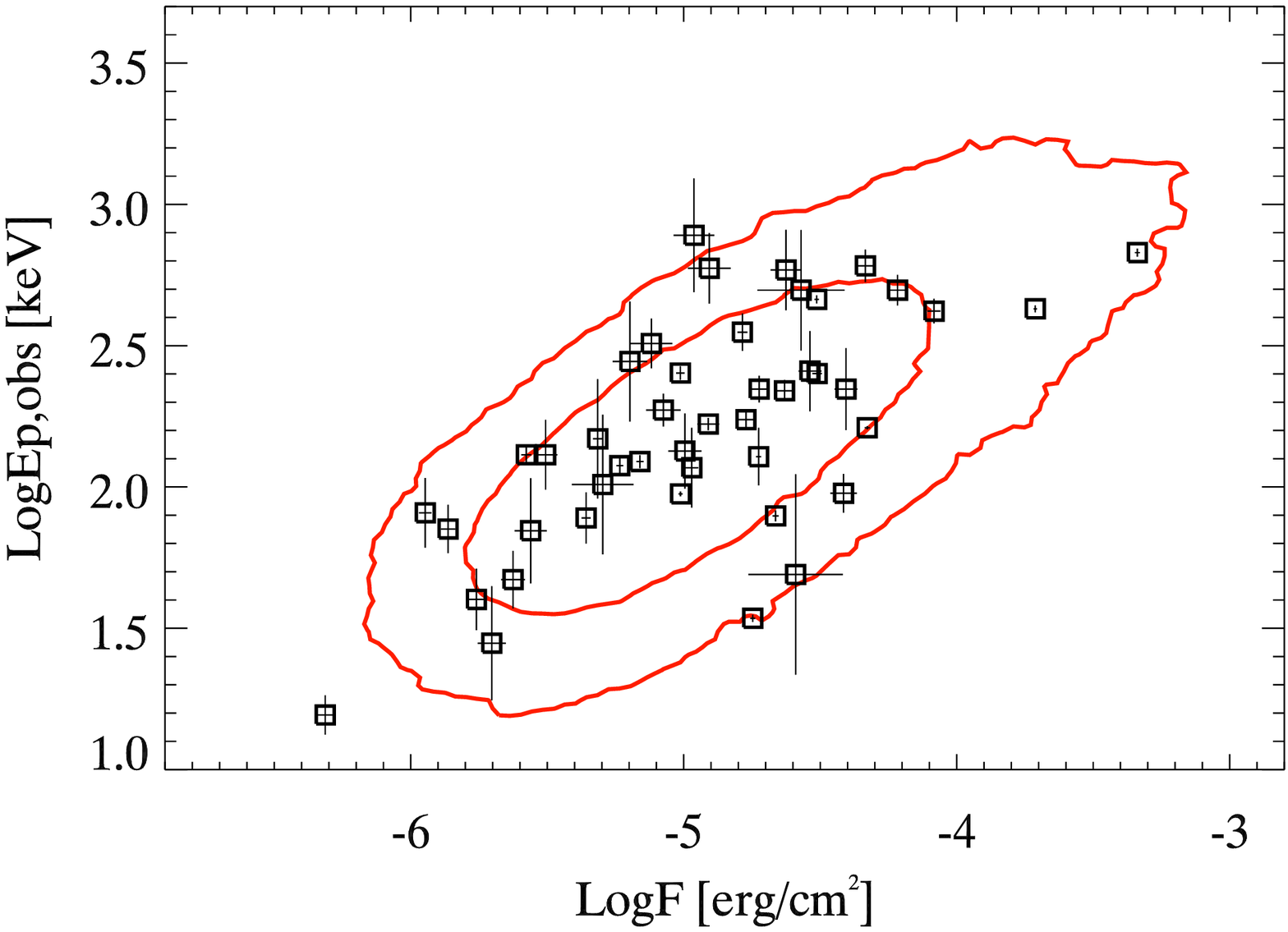} \\
\end{array}$
\end{center}
\vskip -0.6 cm
\caption{
Simulations assuming log--normal distributions of \th\ and \G\ and the
relation $\theta_{\rm jet}^{5/2}$\G=constant.
Left panel: simulated points and real data (black) in the \ama\ plane.
Yellow points are all simulated bursts, blue points are those pointing at us, 
red contours are the distribution of simulated bursts (1 and $\sigma$)
brighter than the peak flux limit of the \sw\ complete sample (i.e. 2.6 ph cm$^{-2}$ s$^{-1}$).
Right panel: Simulated (contours) and real points (black squares) are compared
in the \epo--Fluence observational plane.
}
\label{fg2}
\end{figure*}

%
\begin{table*}
\begin{center}
\begin{tabular}{lllllll}
\hline
\hline
Distrib.  &sample   &$\sigma$        &$\mu$	        &Mode           &Mean           & Median      \\
\hline
\th\      &ALL      &0.916$\pm$0.001 &1.742$\pm$0.002 &2.5$^\circ$   &8.7$^\circ$   &5.7$^\circ$  \\
          &PO       &0.874$\pm$0.010 &3.308$\pm$0.013 &12.7$^\circ$  &40.0$^\circ$  &27.3$^\circ$ \\
          &PO \sw\  &0.527$\pm$0.032 &1.410$\pm$0.043 &3.1$^\circ$   &4.7$^\circ$   &4.1$^\circ$  \\ 
                     
\hline
\G\       &ALL      &1.475$\pm$0.002 &4.525$\pm$0.002 &11    &274     &92     \\
          &PO       &1.452$\pm$0.020 &2.837$\pm$0.025 &2     &49      &17     \\
          &PO \sw\  &0.975$\pm$0.060 &5.398$\pm$0.083 &85    &355     &221    \\         
\hline
\end{tabular}
\caption{
Parameter values ($\mu$ and $\sigma$) obtained by fitting a log--normal function 
to the distributions of \G\ and \th\ (Fig. 3), for 
all the simulated bursts (ALL), for those pointing to us (PO) and for those pointing 
to us and with a peak flux larger than 2.6 ph cm$^{-2}$ s$^{-1}$ (the flux limit of the
complete \sw\ sample) (PO \sw).  
For each distribution are re ported the three moments: the mode, the mean and the median.  
}
\label{tab1}
\end{center}
\end{table*}

\subsection{Results}

\begin{figure}
\hskip -0.5truecm
\includegraphics[width=7.8cm,trim=20 10 20 20,clip=true]{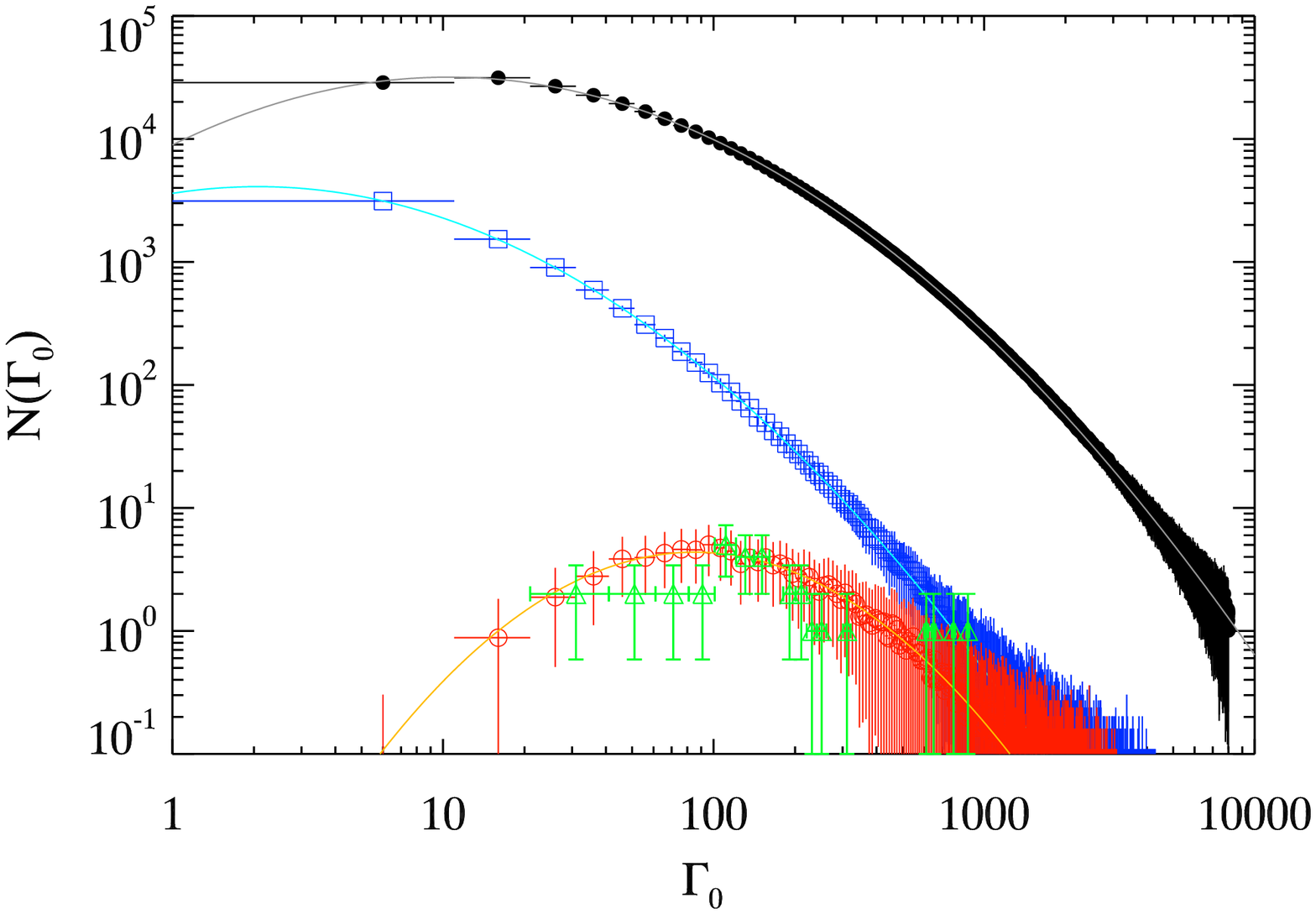}
\includegraphics[width=7.8cm,trim=20 10 20 20,clip=true]{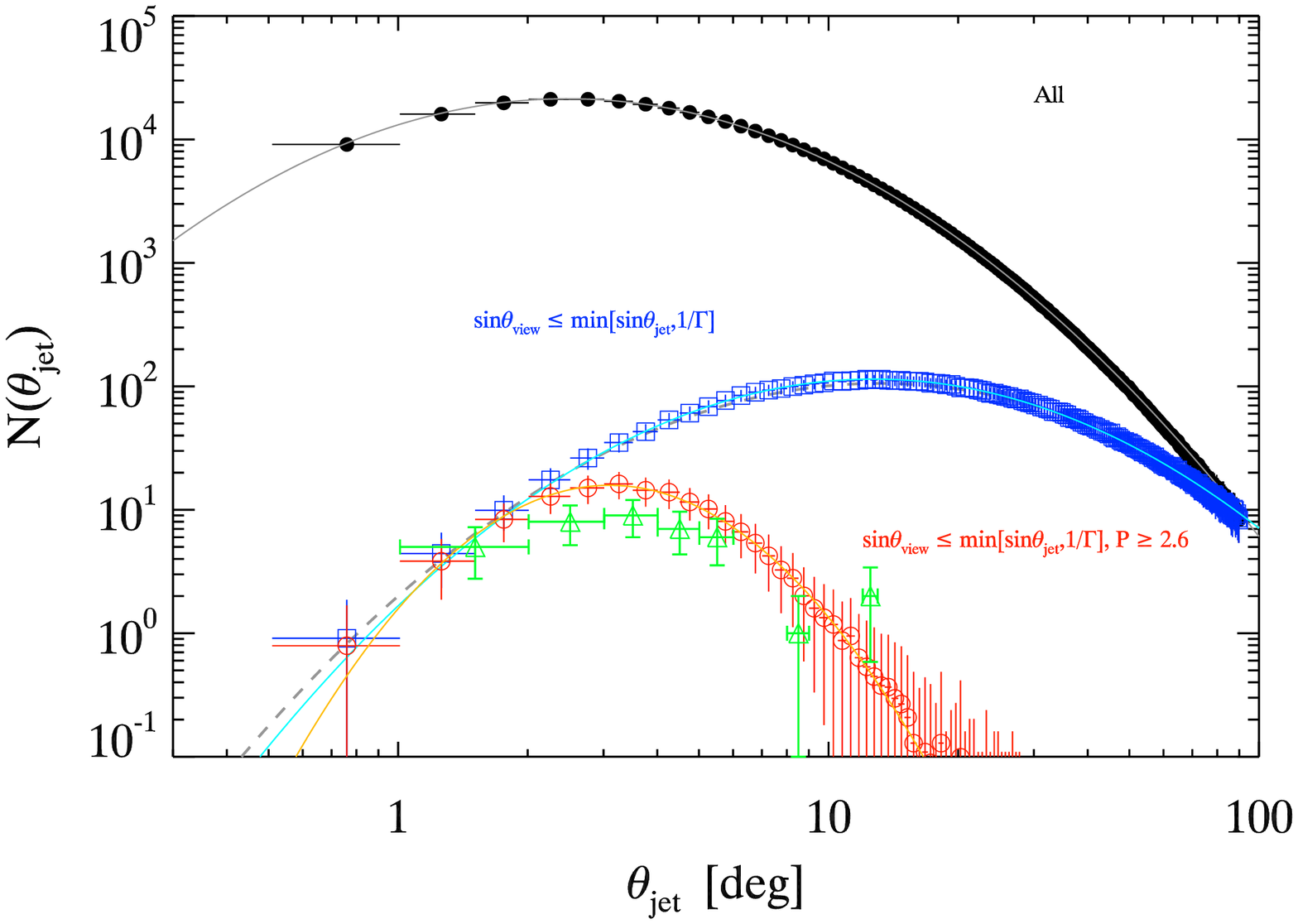}
\vskip -0.2cm
\caption{
Distribution of \G\ (left) and of \th\ (right) of GRBs. 
Black circles: all simulated GRBs;
open blue squares: all simulated GRBs pointing at us; 
open red circles: GRBs pointing at us with
peak flux larger than  2.6 ph cm$^{-2}$ s$^{-1}$ (flux limit of the complete
\sw\ sample); green triangles: 
the $\sim$30 GRBs with \G\ estimated from the onset of the afterglow \cite{ghirlanda2012a}
on the left panel, and the 27 GRBs with measured \th\
collected in \cite{ghirlanda2004}, \cite{ghirlanda2007} on the right panel. 
}
\label{fg3}
\end{figure}

We performed several simulations considering first that 
both \G\ and \th\ have no preferred values, i.e. 
assuming that they are distributed as power--laws,
changing the corresponding slopes.
None of these cases is in agreement with the data.
Then we assumed a broken power law either for \G\ or for \th,
or for both.
For the latter case we do find some agreement, but the distribution
of the simulated points in the \ama\ plane describes a linear correlation,
instead of the observed $E_{\rm p} \propto E_{\rm iso}^{0.6}$.
We then tried log--normal distributions both for \G\ and \th. 
In addition
we assumed that there is a relation between the average values of the two
distributions.
The best results are obtained with $\theta_{\rm jet}^{5/2}$\G=constant (Fig. \ref{fg2}).
Note that the slope of the \ama\ correlation of bright bursts is harder than 
for faint ones (see the blue points in Fig. \ref{fg2}). 
But, curiously, these bright GRBs sample the distribution of the
whole ensemble of bursts (yellow points) better than the fainter ones.
This is because, if we improve our detector sensitivity, we
preferentially see GRBs with larger opening angles. 
This makes
them less energetic and enhances their probability to point at us.
Fig. \ref{fg3} shows (left panel) the distribution of \G\ of
all simulated bursts (black), those pointing at us (blue)
and those (red) that are pointing at us and have a peak flux larger than
2.6 ph cm$^{-2}$ s$^{-1}$ (i.e. the flux limit of the complete \sw\ sample).
The green points correspond to the few GRBs of measured \G\ (left) or \th\ (right).
Tab. \ref{tab1} reports the parameters of the best fitting log--normal
distributions values of \G\ and \th\ for all bursts (ALL), for those pointing at us (PO)
and for those pointing at us with peak flux larger than 2.6 ph cm$^{-2}$ s$^{-1}$ 
(the flux limit of the complete \sw\ sample) (PO \sw).  
 
\section{Conclusions}

The crucial assumption of this study is that all bursts have the same \epcom=1.5 keV and 
\egcom$\sim 2\times 10^{48}$ erg. 
Although there could be a dispersion of these values, our results still hold if the width 
of this dispersion is not larger than the dispersion of the observed quantities. 
The fact that these values are independent of \G\ suggests that the dissipation mechanism
giving rise to the prompt emission is not the transformation of bulk kinetic into
random energy.
If our assumption is true, then the \ghi\ relation is produced by the distribution
of \G\ values, and must be linear (both \ep\ and \egamma\ are proportional to \G).
In turn, the \ama\ relation results from a distribution of jet aperture angles,
with the caveat that, for small values of \G, the radiation collimation angle
is 1$/$\G, not \th.
These bursts will never have a jet--break in the light curve of their afterglow,
and could be mistaken as outliers. 
In our simulations we find that these should be about 6\% of the GRBs pointing at us.
Another important outcome of our study is that we can calculate the fraction of
all GRBs (whether aligned or misaligned) with respect to SN Ibc, 
as a function of redshift.
Taking the recent estimates of the SN Ibc of \cite{grieco2012}, we find that,
locally (i.e. up to $z\sim$1), GRBs are 0.3\% of all SN Ibc.

\end{document}